# Questions and challenges of what powers galactic outflows in active galactic nuclei

Different mechanisms can drive outflows in active galactic nuclei (AGN), but it is often unclear which mechanism dominates, if any. To quantify the impact of AGN feedback on galaxy evolution, the driving mechanism of outflows must be better understood.

**Dominika Wylezalek[1, *] and Raffaella Morganti[2, 3]**

1 European Southern Observatory, Garching, Germany.
2 ASTRON, the Netherlands Institute for Radio Astronomy, Dwingeloo, The Netherlands.
3 Kapteyn Astronomical Institute, University of Groningen, Groningen, The Netherlands.
 * ESO Fellow: dwylezal@eso.org

The regulation of star formation in galaxies and the build-up of their stellar mass are major unresolved issues in extragalactic astronomy. Galactic outflows are an important process that, in conjunction with gas fueling and the efficiency of forming stars, are considered to play a major role in defining how galaxies evolve. The wealth of results from observations with the most advanced facilities have revealed the complexity of the physical conditions of galactic outflows (see other pieces in the Nature Astronomy focus issue). In addition to this, there is still uncertainty and often confusion about whether there is a dominant mechanism driving the outflows and, if so, how to identify it. In galaxies hosting an AGN, the effect of stellar feedback can be outshined by the impact of the energy released by the AGN. This energy release may lead to gas outflows that manifest themselves via two main processes: outflows driven by the radiative energy output of the AGN and outflows related to the mechanical effects of synchrotron-emitting radio jets. The challenge is to identify unbiased diagnostics that help constrain the main driving mechanism of outflows in AGN.

The first step is to separate cases where stellar processes can be the main mechanism. In the case of star-forming galaxies, gas velocities in starburst-driven outflows correlate with the overall star formation rate (SFR) in the galaxy ([1]). Thus, a reliable estimate of the total SFR can be used to estimate if stellar processes contribute significantly to any observed outflows. But even in galaxies with SFRs of several hundreds of $M_{sun}$ yr$^{-1}$, a starburst driven outflow is not expected to exceed velocities of 1000 km s$^{-1}$. For example, 500 km s$^{-1}$ is equivalent to energies of 1 keV per particle and is difficult to achieve with stellar processes ([2]), such that any higher outflow velocities observed may be attributed to an AGN-driven outflow. However, the details are often not easy to disentangle since there is no clear dividing line between AGN- and starburst driven outflows. Recent models even show that stellar-driven outflows that enrich the circumgalactic medium with metals, may actually enhance the effect of AGN outflow and feedback signatures ([3]), showcasing that stellar- and AGN-driven outflows may not only often co-exist but also co-operate.

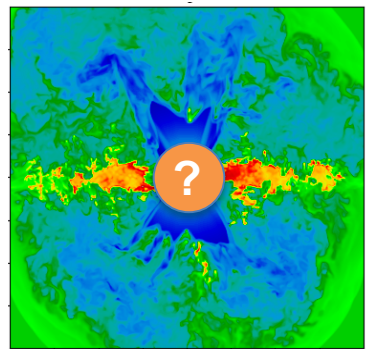

Figure 1: Hydrodynamical simulation of an AGN-driven wide-angle outflow propagating perpendicular to the galaxy disk (adapted from [18]). The gas density (blue: low density, red: high density) is shown here in an edge-on slice through the center of a galaxy. What is the relative importance of different outflow driving mechanisms?

The above conundrum is especially troublesome in low-luminosity AGN (typically $L_{bol}<10^{43}$ erg s$^{-1}$), where the velocities of outflows can be low (v < 500 km s$^{-1}$) and consistent with being driven by either AGN or stellar processes. In these cases, it may be useful to consider a line excitation analysis, e.g. through diagnostics like the Baldwin, Phillips & Terlevich (BPT) diagrams or WHAN (Hα equivalent width vs. [NII]/Hα) analysis (see e.g. [4]). Additionally, in cases of spatially resolved spectroscopic observations, the geometry of the outflow may also help in assessing the dominant driving process. For example, [5] shows that the base of the outflow in a low-luminosity Seyfert galaxy is coincident with the unresolved nucleus, supporting the hypothesis that the AGN is indeed the predominant ionising source of the outflowing gas. However, identifying the effects of AGN feedback in outflows relies in most of the cases on 'just' observing higher velocity (e.g. > 500 km s$^{-1}$) components and an outflow power exceeding that predicted by any central starburst. In bona-fide AGN-driven outflows – often found in high luminosity AGN with powerful outflows – the question about the dominant driving mechanism persists and the impact that different types of AGN can have on the evolution of galaxies needs to be investigated (see Fig. 1). Interestingly, the periods in which the black hole is active can be short and repeat during the life of a galaxy and different types of AGN may have different duty-cycles. Thus, it is crucial to have a census of the role that these different AGN have in the feedback process.

AGN-driven outflows are most commonly considered to be driven by the radiative energy output of the AGN. This is because optically-selected AGN are common and are present in galaxies covering a broad range of masses. For such outflows, the mass outflow rates and outflow sizes appear to scale with the AGN bolometric luminosity ([6]), although this conclusion is still limited by small and heterogeneous samples. However, radiation pressure is not always the dominant driving mechanism. Radio jets have been long known to affect their surrounding medium. The occurrence and impact of radio jets have been revised in the last years, suggesting that their role should not be underestimated.

The reason why powerful radio jets are often considered less relevant for driving outflows is because they can be very collimated structures, thus piercing the ISM without affecting it too much. However, recent simulations have shown that the evolution of the jets in a clumpy medium can be very different from the classical description proposed for powerful radio jets. Particularly, radio jets in their early phase of evolution and of intermediate and low radio power (radio powers at 1.4 GHz $\lesssim 10^{25}$ W Hz$^{-1}$) can have a larger impact on their host galaxies. Interestingly, such jets are present in up to ~ 30% of the massive galaxy population ([7]). In addition, scaling relations derived e.g. from X-ray cavities tell us that the jet power can be about a factor 100 larger than the jet luminosity. Although this relation becomes more uncertain for lower-luminosity AGN (due, for example, to the effect of entrainment of thermal material), it still suggests that the jet power can be significant even in low-power radio sources.

The commonly used classification into radio-loud and radio-quiet AGN often affects whether radiation pressure or the mechanical effect of radio jets is thought to be the main driving mechanism. Despite evidence that this separation is obsolete and can be misleading, it often introduces an unnecessary bias ([8]). In sources which have been classified as radio-quiet, the impact of radio jets is typically not considered as a possible driving mechanism. However, the situation is often not so straightforward. Two well-studied objects can serve as eye-opening examples: Mrk 231 ([9]) and IC 5063 ([10]). These objects are both classified as radio-quiet (radio powers at 1.4GHz $\lesssim 10^{24}$ W Hz$^{-1}$), they have extended radio structures, they both show multi-phase outflows, and they both show the presence of an optical AGN. Nevertheless, two different mechanisms – radiation pressure for Mrk 231 and radio jets for IC 5063 – have been suggested to drive their outflows. These examples – together with a handful of other well studied cases – tell us not to rely only on the radio-loud/radio-quiet classification if we are to infer the driving mechanisms of their outflows.

Further indicative of how careful one has to be in drawing conclusions are the studies of samples of optical AGN selected to be radio quiet where (surprisingly) correlations have been found between the width of the optical emission lines (indicating disturbed gas) and the radio power ([11],[12]). Again, this suggests that radio jets can play a role in driving outflows even in radio-quiet AGN. Interestingly, the presence of such a correlation for low power radio sources has suggested the alternative possibility of the radio emission being a by-product of the shock induced by a radiatively-driven outflow. If this scenario is confirmed, the radio emission of radio-quiet AGN with fast outflows could be a result rather than the driver of a radiatively-driven outflow. Clear cases of such a situation have not been identified yet. A possible candidate (IRAS 17020+4544) for this scenario has been shown to have radio jets (see e.g. [13]) once follow-up deep and high resolution radio images had been obtained. Additionally, precise measurements of the radio spectral index may help in assessing the origin of the radio emission. Thus, we argue that the clear separation between radiative and mechanical driving mechanisms of outflows can still be difficult and both should be considered even for AGN classified as radio-quiet.

Cases where jets likely play the prominent role have been identified mostly in two ways: by using indirect evidence and the energetics (although large uncertainties are associated with energy estimates [see *Perspective Article* by [18] in this issue]). In cases where the bolometric luminosity of the AGN is estimated to be too low and not large enough to drive the outflow, radio jets are considered as an alternative. In many of these cases, radio jets can indeed be energetic enough, even in radio-quiet objects. High resolution radio images and spatially resolved kinematic maps of the outflow allow us to spatially map and compare the radio emission and the outflow. This comparison has been performed in a few sources, e.g. IC 5063 ([10]), 4C12.50 ([15]) and 3C 293 ([16]), and the evidence is strong for small-scale radio jets driving these outflows. Thus, high spatial resolution combined with high sensitivity (to trace the low surface brightness components) are the requirements. The need for sensitivity to map structures with low surface brightness should not be underestimated considering that jets tend to become less collimated and their morphology more complex as we move to low-power radio sources. Furthermore, if this can be obtained at different frequencies, it can provide the spectral information suggested by theoretical models [17] to be able to help disentangling the origin of the radio emission. These requirements are of course time consuming, especially for weak radio sources, but fortunately several high-quality radio surveys are currently or will soon be underway making this task easier.

Disentangling the driving mechanism of gaseous outflows still requires progress in the quality of data available that should include multi-wavelength information and include samples of objects covering broad ranges of optical and radio luminosities. Only then will it be possible to draw strong conclusions about the occurrence of different mechanism driving outflows in different types of AGN.

**Acknowledgements**: We thank the organisers and the participants of the Lorentz Workshop 'Reality and Myths of AGN Feedback' for stimulating discussions that inspired the writing of this comment. We also thank the Lorentz Center for their hospitality. RM gratefully acknowledges support from the European Research Council under the European Union's Seventh Framework Programme (FP/2007-2013) /ERC Advanced Grant RADIOLIFE-320745.